\documentclass[%
aip,
amsmath,amssymb,
reprint,%
]{revtex4-1}

\usepackage{dblfloatfix}
\usepackage{graphicx}
\usepackage{amsmath}
\usepackage{mathtools}
\usepackage[titletoc, toc, page]{appendix}
\usepackage{chemformula}
\usepackage{amssymb}
\usepackage{float}
\usepackage{hyperref}
\usepackage[]{color}
\usepackage{subfig}
\usepackage{afterpage}%
\begin{document}
		\author{Amin Bakhshandeh}
	%	\email{bakhshandeh.amin@gmail.com}
	\affiliation{Instituto de F\'isica, Universidade Federal do Rio Grande do Sul, Caixa Postal 15051, CEP 91501-970, Porto Alegre, RS, Brazil.}

	\author{Yan Levin}
	\email{levin@if.ufrgs.br}
	\affiliation{Instituto de F\'isica, Universidade Federal do Rio Grande do Sul, Caixa Postal 15051, CEP 91501-970, Porto Alegre, RS, Brazil.}
 	\title  
 { Titration in Canonical and Grand-Canonical Ensembles}
	
	\begin{abstract}
		We discuss problems associated with the notion of pH in heterogeneous systems.   For homogeneous systems, standardization  protocols lead to a well defined quantity, which although different from  S\o rensen's original idea of pH, is well reproducible and has become accepted as the measure of the ``hydrogen potential".  On the other hand, for heterogeneous systems, pH defined in terms of the chemical part of the electrochemical activity is thermodynamically inconsistent and runs afoul of the Gibbs-Guggenheim principle that forbids splitting of the electrochemical potential into separate chemical and electrostatic parts  -- since only the sum of two has any thermodynamic meaning.  The problem is particularly relevant for modern simulation methods which involve charge regulation of proteins, polyelectrolytes, nanoparticles, colloidal suspensions etc.  In this paper we show	that titration isotherms calculated using semi-grand canonical simulations can be very different from the ones obtained using canonical reactive Monte Carlo simulations.  
		
	\end{abstract}
	\maketitle
	\section{Introduction}\label{introduction}
	The concept of pH was first introduced by S\o rensen~\cite{sorensen1909erganzung} in  1909.  The original definition referred to pH as $-\log_{10}[{c_{\text{H}^+}}/c^\ominus]$, where $c^\ominus$ =1M is the standard concentration.  Since in practice pH is measured using electrodes, S\o rensen, later redefined pH in terms of activity of hydronium ions, pH$=-\log_{10} \left[a_+/c^\ominus\right]$, which was thought to be related to the electromotive force (EMF) measured by the system of electrodes through the Nernst equation. Later Linderstr\o m-Lang recognized that the  experimental procedure used to measure pH did not lead exactly to pH= $-\log_{10}[{c_{\text{H}^+}}/c^\ominus]$, nor to 
	pH$=- \log_{10} \left[a_+/c^\ominus\right]$, but to some other quantity which due to the convenience, became widely accepted as the measure of the hydrogen potential~\cite{WinNT}. The problem with a direct measurement of pH is that separation of the electrochemical potential into chemical and electric potential is purely arbitrary, since only the sum of two has any physical meaning.  The Gibbs-Guggenheim principle states that the difference of electrostatic potential between two points located in regions of different chemical composition can not be measured~\cite{guggenheim1936lxxxiii,taylor2002electromotive}.  As early as 1899 Gibbs wrote in a letter~\cite{gibbs1961reprinted}:  ``Again, the consideration of
	the electrical potential in the electrolyte, and especially the consideration of the difference of potential in electrolyte and electrode, involves the consideration of quantities of which we have no apparent means of physical measurement, while the difference of potential in pieces of metal of the same kind attached to the electrodes is exactly one of the things which we can and do measure''.   In 1929, Guggenheim~\cite{GuggenheimI} formalized the observation of Gibbs by stating that  ``the decomposition
	of the electrochemical potential, into the sum of a chemical term $\mu$ and an
	electrical term $e \psi$ is quite arbitrary and without physical significance. In
	other words the chemical potential, or the activity of a single ion, and the electric potential difference between two points in 
	different media are conceptions
	without any physical significance. "~\cite{GuggenheimII}
	
	The confusion between exactly what can and is being measured has led to a proliferation of ``local" pH measurements in soft matter and biophysics literature.   The problem has become particularly acute, since the modern simulation methods employed to study charge regulation of protein and polyelectrolyte solutions often rely on constant pH (cpH) algorithms, which are intrinsically semi-grand canonical~\cite{labbez2006new,bakhshandeh2022reactive,comm}.  In such procedure, pH is specified inside a reservoir of acid and salt, and the protonation state of a protein, polyelectrolyte, or colloidal suspension is calculated using a suitably constructed Monte Carlo algorithm that must respect the detailed balance.  Since only microions are exchanged between the simulation box and the reservoir, the two must be at different electrostatic potential.  For an experimental system in which a colloidal suspension is separated from an external reservoir of acid and salt, this is known as the Donnan potential.  Traditionally pH is defined in terms of the chemical part of the electrochemical potential.  However, since Gibbs-Guggenheim principle forbids us from breaking up the electrochemical potential into separate electrostatic and chemical contributions, such definition appears to be thermodynamically unacceptable.

	In practice pH is measured using EMF between a glass
	or hydrogen electrode and a saturated calomel (reference) electrode. Consider a colloidal suspension separated 
	from a reservoir by a semipermiable membrane that allows free movement of ions, but restricts colloidal particles to system's interior, see Fig. 1.  If the calomel reference electrode is placed in the reservoir and the EMF is measured between it and the hydrogen electrode, one finds constant EMF  independent of the position of the hydrogen electrode, either in the reservoir or in the system's interior,  {\color{black} {see Fig. 1 panels (a) and (b)}}.  This is a clear indication of a constant electrochemical potential of hydronium ions 
	across both system and the reservoir.  On the other hand, if the two electrodes are placed inside the colloidal suspension, the EMF will depend on the distance of the reference electrode from the membrane, {\color{black} {see Fig. 1 panels (c)}}.  Clearly, such measurement would result in thermodynamically ill-defined ``local" pH.   Such experimental measurements were performed by Teorell et.{\it{al}}~\cite{craxford1938xcv} more than 85 years ago.   Already at that time he noted the difficulties with the usual definition of pH when applied to heterogeneous systems. One can argue that if both electrodes are placed deep into the system, far away from  membrane, the resulting EMF will stabilize and will allow us to define the system pH, which will be different from that of the reservoir.  This is correct, but does not resolve the underlying problem  arising from the violation of the Gibbs-Guggenheim principle~\cite{craxford1938xcv}.  For example, consider now a colloidal suspension in a  gravitational field~\cite{van2003defying,philipse2004remarks,warren2004electrifying}.
	Because of finite buoyant mass, colloidal column will become progressively rarefied {\color{black} {with the height   -- characteristic gravitational length of colloidal particles is between micrometers and millimeters.}} On the other hand {\color{black} {on experimental length scale, ionic buoyant mass is negligible}}. Therefore, the top part of suspension will be composed of a pure acid-salt electrolyte, with a well defined pH, since according to the Gibbs-Guggenheim principle, this is the region of uniform chemical composition in which one can measure the electrostatic potential difference between two points.  In the present case, the gravitational field plays the role of a membrane that establishes inhomogeneity of suspension.  This results in a height dependent Donnan potential $\varphi_D(z)$ along the column, which in turn leads to different ionic concentrations at each $z$.  Nevertheless, if we place our reference electrode in the top (colloid-free) portion of the suspension, we will get exactly the same EMF (and consequently the same pH) independent of the placement of the hydrogen electrode inside the colloidal column.  On the other hand, if  the reference electrode is moved into colloid-dense region, each different position will lead to a 
	different EMFs and, consequently, a different pH.     One might argue that if both hydrogen and calomel electrodes are placed at {\it exactly} the same height $z$, the pH obtained using such measurement will have some physical meaning.  Such proposition, however, once again seems untenable in view of the Gibbs-Guggenheim principle, since only the full electrochemical potential has any thermodynamic meaning. The confusion in literature is such that in a paper published some years back 
	Brezinski wrote: ``the uncertainty regarding interpretation of pH readings for
	colloids has led to the opinion that the pH value of
	neither the sediment nor the supernatant is very
	meaningful or useful for characterizing colloids"~\cite{BREZINSKI1983347}.  Based on the preceding discussion, such view seems overly pessimistic.  While pH in the homogeneous supernatant of suspension is well defined thermodynamically, in the interior of a highly inhomogeneous suspension it runs afoul of the Gibbs-Guggenheim principle. On the other hand, from a purely theoretical perspective, knowledge of pH in the inhomogeneous part of suspension is completely irrelevant.  Specification of pH and salt concentration in the {\it homogeneous} reservoir (supernatant) should be sufficient to calculate the state of protonation of colloidal particles and their density profile, both of each are easily accessible to experimental measurements.  In theory  -- or simulation -- one could even calculate hydronium density profile inside an inhomogeneous suspension, there is however no clear connection between this local density of hydronium ions and the extra-thermodynamic quantity such as ``local" pH of an inhomogeneous suspension~\cite{warren2004electrifying}.

	When performing classical charge regulation simulations, one has two options  -- either a {\it semi-grand canonical} constant pH (cpH) simulation in which the system is placed in contact with an implicit reservoir or acid and salt~\cite{labbez2006new,bakhshandeh2022reactive}, or a {\it canonical} simulation in which a fixed number of polyelectrolytes, protons, ions, water molecules are placed inside a simulation box~\cite{barr2012grand,Levincomment}.  The two approaches are very different, requiring  distinct implementations of the Monte Carlo algorithm to take into account protonation/deprotonation moves~\cite{Levincomment}.  When performing cpH simulations, insertion of a proton into the system is accompanied by a simultaneous insertion of an anion, to preserve the overall charge neutrality.  On the other hand,  in a canonical simulation, a proton is transferred from a hydronium molecule inside the simulation box, to a polyelectrolyte monomer, so that the charge neutrality is always preserved.  This requires a completely different implementation of the MC algorithm.  Furthermore, in a canonical simulation pH is not an input parameter, and can only be calculated {\it a posteriori} after the system has equilibrated.  The consistency between the two simulation methods can be tested {\it a posterior}.  For example,
	we can run a cpH simulation, for a given pH and salt concentration in the reservoir.  This will provide us with the average number of protonated groups on polyelectrolytes, as well as with the average number of ions of each type inside the simulation cell.  We can then isolate the system from the reservoir (canonical ensemble) keeping exactly the same number of ions inside the simulation cell as the averages obtained in cpH simulation.  We then strip all the associated protons from polyelectrolyte and place them randomly (in the form of hydronium ions) together with the other ions  into the simulation cell.  We can then run a canonical reactive MC algorithm.  Equivalence between ensembles then requires that we obtain {\it exactly} the same number of protonated groups as was previously found using cpH simulation.  This is precisely what is observed, showing consistency of the two simulation methods~\cite{Levincomment}.
	
	The cpH simulations start with a specified value of pH$_{gc}$ and salt concentration inside the reservoir. 
	On the other hand, in canonical simulations pH$_{c}$ has to be determined  
	{\it a posterior} using Widom insertion method.  If we define pH in the semi-grand canonical system in terms of the total electrochemical potential -- corresponding to keeping the calomel electrode inside the reservoir, while the hydrogen electrode is ``placed" into the simulation cell  --  then the system pH$_{sys}$  will be the same as of the reservoir pH$_{gc}$, and will, in general, be different from pH$_c$ in the canonical system.  On the other hand, if we disregard the Gibbs-Guggenheim principle and separate the Donnan potential from the rest of the electrostatic potential, then the pH inside the system will be {\it different} from  pH$_{gc}$ and the {\it same} as canonical pH$_c$.  This situation corresponds to ``placing" both hydrogen and reference electrode inside the simulation cell of a semi-grand canonical system.  In practice, a calculation of the electrochemical potential in a canonical simulation is quite complicated,  in particular if pH is large,  since the simulation box will have only very few hydronium ions, resulting in very poor statistics.  This led to popularization of thermodynamically poorly defined ``local" pH$({\bf r})=-\log_{10}[c_{H^+}(r)/c^\ominus]$~\cite{landsgesell2019simulations}.    To avoid these difficulties, and clearly demonstrate the effect of ensembles on titration isotherms, in this paper we will use a recently developed theory, which was shown to be in excellent agreement with the explicit ions cpH simulations~\cite{bkh22}.

	\begin{figure}[H]
		\centering
		\includegraphics[width=0.6\linewidth]{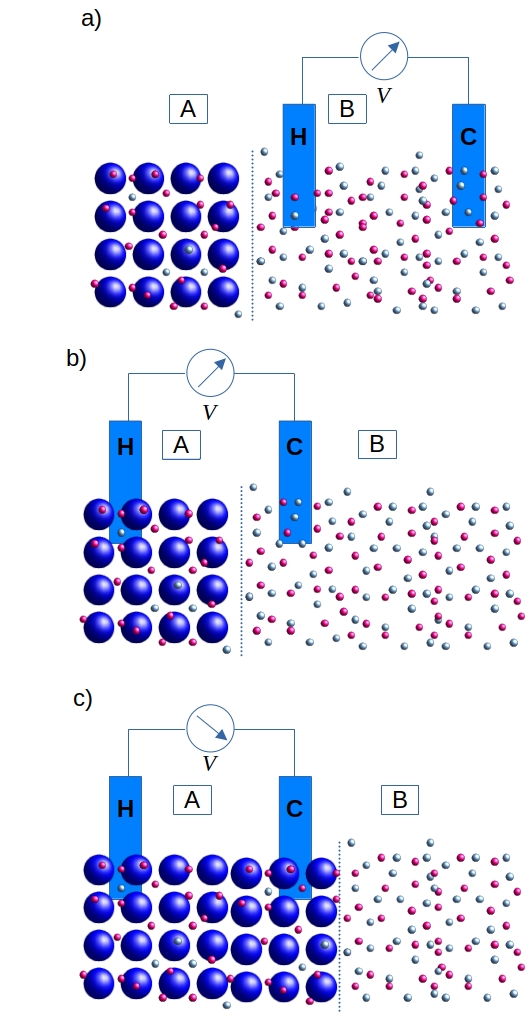}
		\caption{Colloidal crystal separated from a reservoir of acid and salt by a semi-permeable membrane.   {\color{black} Panels (a), (b) and (c) show the different locations of the (reference)  calomel (C)  electrode  and the hydrogen (H) electrode. Note that EMF readings in panels (a) and (b) are the same, while in the panel (c) it is different.}  }
		\label{it1}
	\end{figure}

	\begin{figure}[H]
		\centering
		\includegraphics[width=0.4\linewidth]{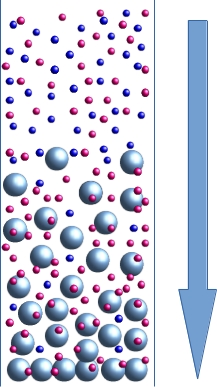}
		\caption{Colloidal suspension in a gravitational field. The top portion is a homogeneous, colloid free electrolyte solution, where pH has a well defined thermodynamic meaning.}
		\label{ it2}
	\end{figure}

	\section{Theory}
	\subsection{Semi-Grand Canonical Titration Theory} 
	To explore the difference between canonical and grand canonical titration, we will use a cell model first introduced by S. Lifson and A. Katchalsky, and  R. A. Marcus~\cite{fuoss1951potential,marcus1955calculation} to study polyelectrolyte and colloidal systems of finite volume fractions.  The model consists of a colloidal particle of radius $a=60$ \AA, placed at the center $r=0$ of a spherical cell of radius R, which is determined  by the volume fraction of the colloidal suspension $\eta_c=a^3/R^3$.    The cell is assumed to be in contact with a reservoir of acid and 1:1 salt at concentrations $c_a$ and $c_s$, respectively.  All ions are treated as hard spheres of diameter $d=4$ \AA\, with a point charge located at the center.  The nanoparticle  has $Z=600$ carboxylic groups of $\text{pK}_a=5.4$, uniformly distributed over its surface.  Ref.~\cite{bkh22} showed that the average number of deprotonated groups of a colloidal particle is given by:
	%%%%%%%%%%%%%%%%%%%%%%%%%%%%%%%%%%%%%%%%%
	\begin{equation} \label{zeff2}
		Z_{eff}=\frac{Z}{1+ 10^{-\text{pH}_{gc}+\text{pK}_a } e^{-\beta(q \varphi_0+\varphi_{disc} -\mu_{sol})}},
	\end{equation} 
	where $q$ is the proton charge.  The pH in the reservoir is determined by : $\text{pH}_{gc}=-\log_{10} \left[a_{\text{H}^+}/c^\ominus\right]$, with the activity of hydronium  ions in the reservoir $a_{\text{H}^+}=c_{\text{H}^+} \exp(\beta \mu_{ex} )$, where $\mu_{ex} =\mu_{CS}+\mu_{MSA}$ is the excess chemical potential.  The non-ideality effects due to Coulomb interactions are taken into account at the mean spherical approximation (MSA) level, while the hard core contribution is calculated using the  Carnahan-Starling equation of state~\cite{carnahan1969equation,carnahan1970thermodynamic,adams1974chemical,MACIEL2018,ho1988mean,ho2003interfacial,levin1996criticality,waisman1972mean,blum1975mean,levin2002electrostatic}:
	\begin{equation}~\label{msa}
		\beta \mu_{MSA} = \frac{\lambda_B\left( \sqrt{1+2 \kappa d}-\kappa d -1\right)} {d^2\kappa},
		\qquad
		\beta\mu_{CS} = \frac{8\eta-9 \eta^2+3\eta^3}{\left(1-\eta\right)^3},
		\qquad
	\end{equation}   
	where  $\eta=\frac{\pi d^3}{3} c_t$,   $c_t=c_s+c_a$ is the total concentration of salt and acid, $\lambda_B=q^2/\epsilon_w k_B T=7.2$ \AA\ is the Bjerrum length, and  $\kappa= \sqrt{8 \pi \lambda_B c_t}$ is the inverse Debye length. The surface groups are characterized by  $\text{pK}_a=-\log_{10}[\text{K}_a/c^\ominus]$, where $\text{K}_a$ is the acid dissociation constant of surface groups and  $ \varphi_0 $ is the mean-field electrostatic potential at the surface titration sites. The ion concentration inside the cell, for  $r \ge a+d/2$, is given by the Boltzmann distribution
	\begin{equation}\label{bolt}
		\rho_i({\bf r}) = c_i e^{-\beta q_i \varphi({\bf r})} \,,
	\end{equation} 
	where $c_i$ is concentration of ions of type $i$ in the reservoir.   
	The mean field potential, $\varphi(r)$, satisfies the Poisson-Boltzmann equation for $r \ge a+d/2$,
	\begin{equation}\label{mpb}
		\nabla^2 \varphi(r) =  \frac{8 \pi q }{\epsilon_w} \left(c_a+c_s\right) \sinh[\beta q \varphi(r)] ,
	\end{equation} 
	and Poisson equations for $a<r<a+d/2$.   The  discreteness of surface sites is taken into account 
	self consistently using the electrostatic potential~\cite{bkh22} 	
	\begin{equation}\label{EOCP2}
		\beta \varphi_{disc}    = -\frac{\lambda_B M Z_{eff}}{ a \sqrt{Z}}, 
	\end{equation}
	%%%%%%%%%%%%%%%%%%%%%%%%%%%%%%%%%%%%%%%%%%
	where $M$ is the Madelung constant of the two dimensional One Component Plasma (OCP) in a hexagonal crystal state~\cite{bkh22,bakhshandeh2019}.  
	Finally,  $ \mu_{sol}$ is the  electrostatic solvation free energy of an isolated charged site:
	\begin{equation}\label{sol}
		\beta \mu_{sol}=\frac{\lambda_B}{2}\int_0^\infty\frac{k-\sqrt{\kappa^2 +k^2}}{k+\sqrt{\kappa^2 +k^2}} e^{- k d }dk.\ 
	\end{equation}
	Solving numerically the non-linear PB equation with the boundary condition of vanishing electric field at the cell boundary (charge neutrality) and colloidal charge determined self-consistently by Eq. (\ref{zeff2}), we obtain the number of protonated groups for a given pH$_{gc}$. Note that at the surface of the cell there is a jump in the electrostatic  potential  -- the reservoir is taken to be at zero potential, while at the cell boundary, $r=R$, the electrostatic potential is calculated to have a finite value $\varphi_D$.   
	This is the Donnan potential of a suspension that is in contact with a reservoir of acid and salt through a semi-permeable membrane.  The titration curves, for a fixed concentration of 1:1 salt inside the reservoir are presented by dashed red lines in Figs. 1 and 2 as a function of pH  {\it in the reservoir}. 
	
	\subsection{  Canonical Titration Theory} 
	
	Suppose we run a cpH simulation from which we calculate the average number of deprotonated groups $Z_{eff}$, the average number of free hydronium ions, and the average number of sodium and chloride ions inside the cell.  We then isolate the cell from the reservoir (canonical ensemble), keeping exactly this number of free ions inside the cell and fixing the colloidal charge at $Q=-q Z_{eff}$.  Since the cell is no longer connected with the external reservoir, there is no Donnan potential at the cell boundary, and the 
	electrostatic potential must be continuous between inside and outside the cell.  Since outside the cell $\phi=0$, we conclude that 
	$\phi(R)=0$. 
	
	At the  level of approximation of the present theory, the distribution of hydronium ions inside the cell of a {\it canonical systems} is given by:
	\begin{equation}~\label{eqt}
		\rho_{{\rm H}^+}\left(\bf{r}\right)=\frac{N_{{\rm H}^+} \mathrm{e}^{-\beta q \phi(\bf{r})}}{4 \pi \int_{a+\frac{d}{2}}^R r^2  dr \mathrm{e}^{-\beta q \phi(\bf{r})}}
	\end{equation}
	where $N_{{\rm H}^+}$ is the number of free hydronium ions inside the cell and $\phi(r)$ is the mean field electrostatic potential.  
	
	The electrochemical  potential of hydroniums inside the cell is:
	\begin{equation}~\label{mu}
		\beta\mu_{c} =\ln\left[\rho_{{\rm H}^+}({\bf r}) \right] +\beta q \phi({\bf r}) + \mu_{ex}
	\end{equation}
	where $\mu_{ex}$ is the excess chemical potential due to the electrostatic and steric interactions between the ions, which at the level of the present theory we take to be constant and equivalent to  $\mu_{CS}+\mu_{MSA}$ in the reservoir. Clearly, the fact that the system became disconnected from the reservoir after equilibration does not affect the distribution of hydronium ions inside the cell, which must remain exactly the same as before.  The only difference is that the canonical electrostatic 
	potential is shifted from its grand canonical value by the Donnan potential  $\phi(r)=\varphi(r)-\varphi_D$, which does not affect the distribution given by Eq.(\ref{eqt}).  Therefore,  the hydronium density profile, Eq.(\ref{eqt}), can also be written in terms of the acid  concentration $c_a$ in the original reservoir, see Eq. (\ref{bolt}), and the Donnan potential:
	%%%%%%%%%%%%%%%%%
	\begin{equation}~\label{eqt2}
		\rho_{{\rm H}^+}\left(\bf{r}\right)=c_a \mathrm{e}^{-\beta q \phi({\bf r})-\beta q \varphi_D}.
	\end{equation}
	%%%%%%%%%% 
	Substituting this expression into Eq. (\ref{mu}), we obtain the relation between canonical and 
	semi-grand canonical electrochemical potentials:
	\begin{equation}~\label{eqt1}
		\mu_{c}  = \mu_{gc}- q \varphi_D.  
	\end{equation}
	The activity of hydronium ions inside an isolated suspension is then $a_{H^+}=\exp[ \beta \mu_c]/c^\ominus$, so  that canonical and semi-grand canonical pH are found to be related by:
	\begin{equation}\label{key}
		\text{pH}_c = \text{pH}_{gc} + \frac{\beta q \varphi_D}{\ln10}
	\end{equation}

	%if we put Eq.~\ref{eqt1} equal to chemical potential reservoir Eq.~\ref{msa} one can get Eq.~\ref{key} 
	%As a result we have
	%	\begin{equation}~\label{eqt1}
		%	\beta	\mu_{ex} + \ln a_+ +\mu_{corr}   =\ln\left(\frac{C_+ \exp(-\beta q \varphi_D)}{C^\ominus}\right) +\mu_{corr}
		%\end{equation}
		\section{Results and discussion}
		
\begin{figure}
			\centering
			\includegraphics[width=0.7\linewidth]{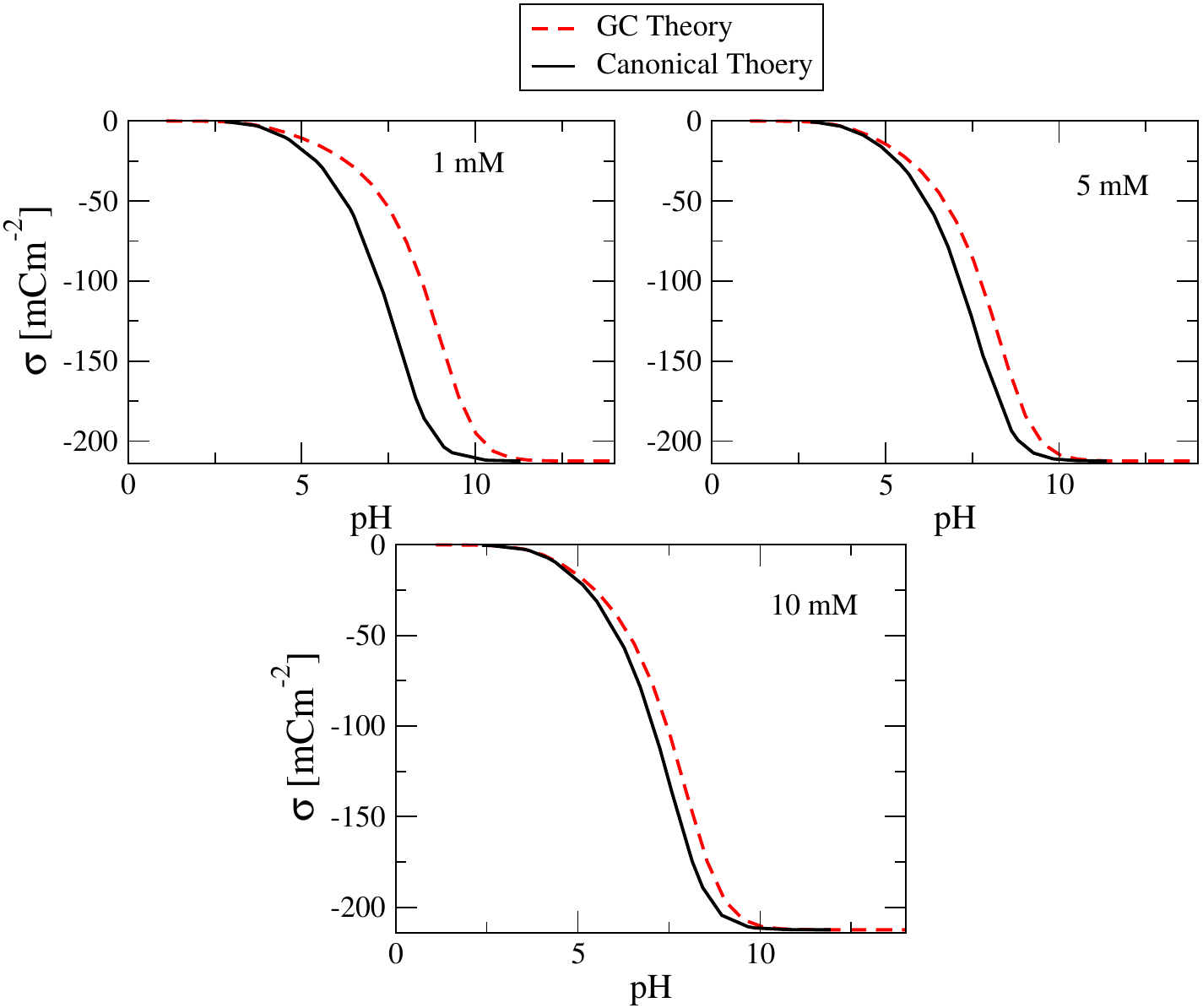}
			\caption{The titration isotherms for canonical and semi-grand canonical ensembles.  For   an open system the pH refers to the pH in the reservoir, while for a closed system it is for the interior of suspension.  Similarly for an open system $c_s$ refers to salt content in the reservoir, while for a closed system it is salt concentration inside the system.  
				At higher salt concentrations the difference between canonical and semi-grand canonical titration curves vanishes. {\color{black} Colloidal volume fraction is $\eta=11\%$, in all cases.  Colloidal surface charge density $\sigma$ is measured in millicoulombs per m$^2$.}}
			\label{fig1}
		\end{figure}

\begin{figure}
			\centering
			\includegraphics[width=0.7\linewidth]{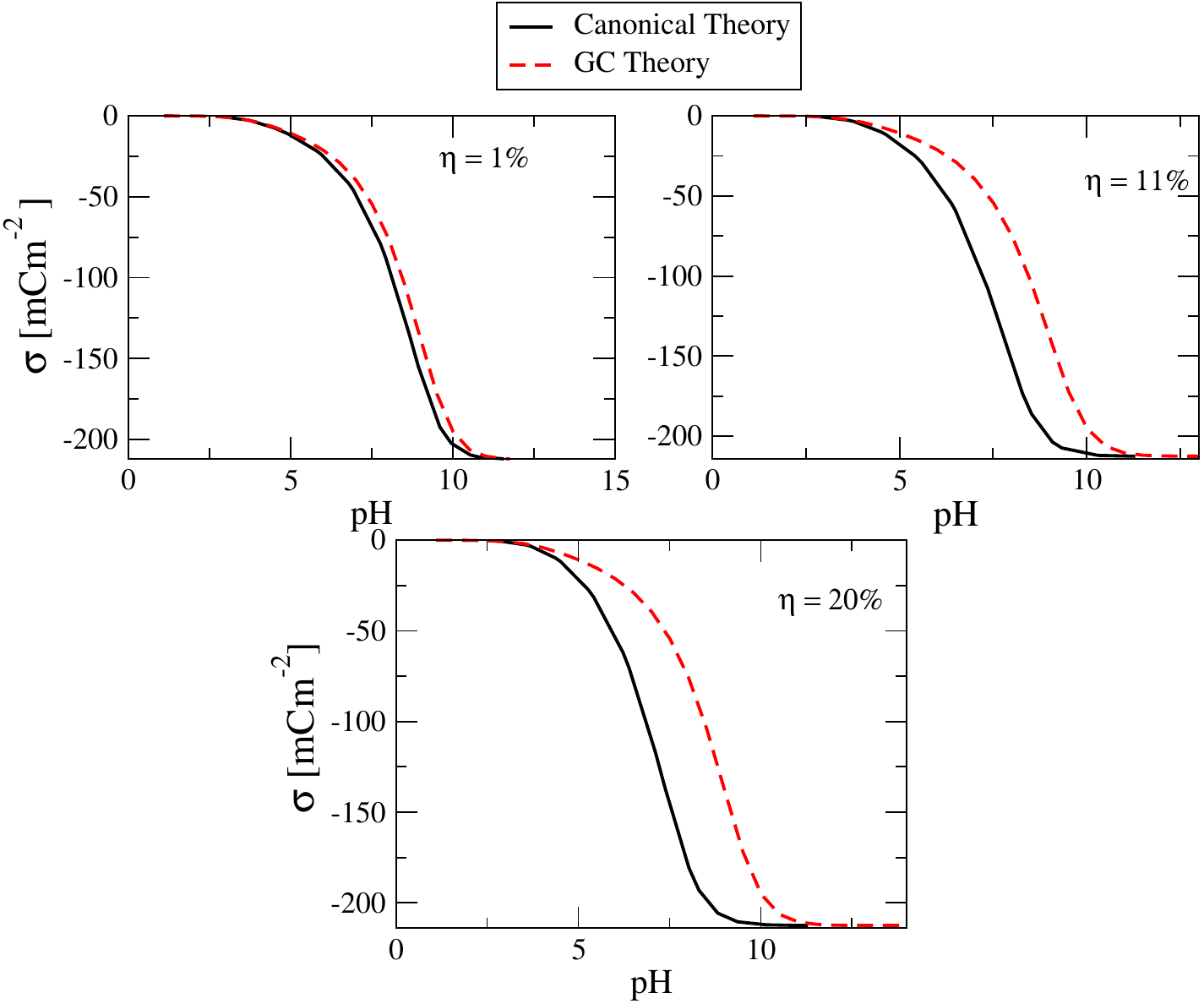}
			\caption{The titration isotherms for canonical and semi-grand-canonical systems of 
				different colloidal volume fractions $\eta$.  Salt concentration is $c_s=1$ mM.  For dilute suspensions (low colloidal volume fraction) the difference between ensembles vanishes. {\color{black} Colloidal surface charge density $\sigma$ is measured in millicoulombs per m$^2$.}}
			\label{fig2}
		\end{figure}

		In Figs. \ref{fig1} and \ref{fig2} we present the titration isotherms for colloidal suspensions of various volume fractions and salt concentrations.  The red dashed curves correspond to systems in which colloidal particles are separated from the acid-salt reservoir of pH$_{gc}$ and $c_s$, by  a semi-permeable membrane.  On the other hand, the black solid curves correspond to titrations performed in  isolated colloidal suspensions containing  a fixed salt concentration $c_s$, indicated in the figures.  To calculate these canonical titration curves, the concentration of salt in the reservoir is adjusted to get the desired concentration of salt inside the system, while solving the PB equation with the boundary conditions of vanishing electric field at $r=R$ and the nanoparticle charge determined by  Eq.  (\ref{zeff2}).  The pH$_c$ is then obtained using equation (\ref{key}).  We see that for suspensions of high volume fractions and low salt content, the canonical titration curves are very different from their semi-grand canonical counterparts.  If salt content of suspension increases or if the volume fraction of colloidal particles decreases, we see that the difference between titration isotherms vanishes.  This explains why these problems were not previously observed in cpH simulations used to study biologically relevant proteins.  Such simulations are usually conducted at physiological concentrations of electrolyte, when the difference between canonical and grand canonical pH vanishes.
		%%%%%%%%%%%%%%%
		%%%%%%%%%%%%%%%%%%	 	  	 	 
		
		\section{Conclusions} 
		
		After 115 years  the measure of ``hydrogen potential"  still causes 
		conceptual and practical difficulties.  The original 
		idea of S\o rensen was to relate pH directly to the concentration of hydronium ions.  
		The complexity of measuring local concentrations  of protons has led him to later redefine pH in terms of activity of hydronium ions.  This, however, resulted in a whole new set of difficulties, since the measurements of individual activity coefficients are not possible by any electrochemical means.   What is in fact being measured using hydrogen and calomel electrodes,  
		is neither activity nor concentration of hydronium ions, but some other quantity~\cite{bjerrum1958electrical}.  Nevertheless, due to standardization of such measurements, they have become well accepted by the scientific community~\cite{WinNT}.  For homogeneous, single phase systems, the situation, therefore, appears to be reasonably well understood.  The difficulties arise when one tries to extend such 
		measurements to heterogeneous systems, such as colloidal suspensions in gravitational fields, or even colloidal lattices in which translational symmetry is broken, and  the electrostatic potential is a strongly inhomogeneous function of position.  The Gibbs-Guggenheim principle forbids us from splitting the electrochemical potential into separate chemical and electrostatic parts, since only the sum of two has any thermodynamic meaning.~\footnote { For reasons why it is not possible to uniquely define the ``mean" electrostatic potential see the discussion of N. Bjerrum in Ref. \cite{bjerrum1958electrical}}
		This suggests that the correct definition of activity should involve the full electrochemical potential $a_{H^+}=\exp[\beta \mu]/c^\ominus$,  where $\mu=\mu_{chem}({\bf r})+q \varphi({\bf r})$.   Although both the chemical potential $\mu_{chem}(\bf r)$ and the {\it total} electrostatic potential $\varphi({\bf r})$ are local functions of position, their sum is constant throughout an inhomogeneous system in which protons are in equilibrium.  Such definition, however, would make activity of protons  --  and  pH  -- in heterogeneous systems with 
		Donnan equilibrium to be the  same on both sides of a semi-permeable membrane transparent to H$^+$, even though the concentrations of 
		hydronium ions on the two sides of such membrane are  different.  The price of such definition would, therefore, be to  move the notion of pH even farther from S\o rensen's original idea of measuring local hydronium concentration.  The gain, however, would be to make pH a true thermodynamic variable, directly related to the electrochemical potential .   
		The current state of affairs seems to be untenable for heterogeneous systems  
		in which the local electrostatic potential is a strongly inhomogeneous function of position.  It is unclear what thermodynamically relevant information  can be extract from pH measurements that are based on standard protocols, 
		in which position of both reference calomel and hydrogen electrodes is changed throughout the measurements.  The notion of  ``local" activity lacks any thermodynamic meaning and cannot be measured.  If such measurements are attempted, the results will be ``accidentally determined" by the sample preparation and electrode possitionining, as was  already stressed by Guggenheim almost 100 years ago~\cite{warren2004electrifying}.  This explains the confused state of affairs in colloidal science exemplified by the, so called, ``suspension effect" in which the pH measured in a charged sediment is found to be very different from that of the supernatant~\cite{BREZINSKI1983347}.  				
		In the present paper,  we suggest that for strongly inhomogeneous systems only pH$_{gc}$ in the homogeneous part has  any significance.  If one wants to study the thermodynamics  -- and statistical mechanics -- of inhomogeneous suspensions  -- such as for example density profiles of colloidal particles in a gravitational field -- only pH in the supernatant can be used as an input in any theoretical investigations, the pH in the sediment will be ``accidental"~\cite{warren2004electrifying}.  Specification, of pH$_{gc}$  in the ``reservoir" avoids the difficulties associated with splitting the electrochemical potential into separate chemical and electrostatic contributions.  A corollary of this is that titration curves in open systems that are in contact with a reservoir  -- plotted as a function of  pH$_{gc}$  -- can be significantly shifted  from titration curves calculated  for closed (canonical) systems of exactly the same volume fraction and electrolyte concentration, as is demonstrated in the present paper.    
		\section{Note}
		The authors declare no competing financial interest.
		\section{Acknowledgments}
		
		YL would like to acknowledge interesting conversations with Christian Holm and Peter Ko\v{s}ovan.  This work was partially supported by the CNPq, the CAPES, and the National Institute of Science and Technology Complex Fluids INCT-FCx.

		\bibliography{ref}

\begin{thebibliography}{34}%
\makeatletter
\providecommand \@ifxundefined [1]{%
 \@ifx{#1\undefined}
}%
\providecommand \@ifnum [1]{%
 \ifnum #1\expandafter \@firstoftwo
 \else \expandafter \@secondoftwo
 \fi
}%
\providecommand \@ifx [1]{%
 \ifx #1\expandafter \@firstoftwo
 \else \expandafter \@secondoftwo
 \fi
}%
\providecommand \natexlab [1]{#1}%
\providecommand \enquote  [1]{``#1''}%
\providecommand \bibnamefont  [1]{#1}%
\providecommand \bibfnamefont [1]{#1}%
\providecommand \citenamefont [1]{#1}%
\providecommand \href@noop [0]{\@secondoftwo}%
\providecommand \href [0]{\begingroup \@sanitize@url \@href}%
\providecommand \@href[1]{\@@startlink{#1}\@@href}%
\providecommand \@@href[1]{\endgroup#1\@@endlink}%
\providecommand \@sanitize@url [0]{\catcode `\\12\catcode `\$12\catcode
  `\&12\catcode `\#12\catcode `\^12\catcode `\_12\catcode `\%12\relax}%
\providecommand \@@startlink[1]{}%
\providecommand \@@endlink[0]{}%
\providecommand \url  [0]{\begingroup\@sanitize@url \@url }%
\providecommand \@url [1]{\endgroup\@href {#1}{\urlprefix }}%
\providecommand \urlprefix  [0]{URL }%
\providecommand \Eprint [0]{\href }%
\providecommand \doibase [0]{http://dx.doi.org/}%
\providecommand \selectlanguage [0]{\@gobble}%
\providecommand \bibinfo  [0]{\@secondoftwo}%
\providecommand \bibfield  [0]{\@secondoftwo}%
\providecommand \translation [1]{[#1]}%
\providecommand \BibitemOpen [0]{}%
\providecommand \bibitemStop [0]{}%
\providecommand \bibitemNoStop [0]{.\EOS\space}%
\providecommand \EOS [0]{\spacefactor3000\relax}%
\providecommand \BibitemShut  [1]{\csname bibitem#1\endcsname}%
\let\auto@bib@innerbib\@empty
%</preamble>
\bibitem [{\citenamefont {S{\"o}rensen}(1909)}]{sorensen1909erganzung}%
  \BibitemOpen
  \bibfield  {author} {\bibinfo {author} {\bibfnamefont {S.}~\bibnamefont
  {S{\"o}rensen}},\ }\bibfield  {title} {\enquote {\bibinfo {title}
  {Erg{\"a}nzung zu der abhandlung: Enzymstudien ii: {\"U}ber die messung und
  die bedeutung der wasserstoffionenkonzentration bei enzymatischen
  prozessen},}\ }\href@noop {} {\bibfield  {journal} {\bibinfo  {journal}
  {Biochem. Z.}\ }\textbf {\bibinfo {volume} {22}},\ \bibinfo {pages}
  {352--356} (\bibinfo {year} {1909})}\BibitemShut {NoStop}%
\bibitem [{Win(2010)}]{WinNT}%
  \BibitemOpen
  \href {https://publications.iupac.org/ci/2010/3202/1_mfcamoes.html} {\enquote
  {\bibinfo {title} {A century of ph measurement},}\ } (\bibinfo {year}
  {2010})\BibitemShut {NoStop}%
\bibitem [{\citenamefont {Guggenheim}(1936)}]{guggenheim1936lxxxiii}%
  \BibitemOpen
  \bibfield  {author} {\bibinfo {author} {\bibfnamefont {E.}~\bibnamefont
  {Guggenheim}},\ }\bibfield  {title} {\enquote {\bibinfo {title} {Lxxxiii. on
  the meaning of diffusion potential},}\ }\href@noop {} {\bibfield  {journal}
  {\bibinfo  {journal} {The London, Edinburgh, and Dublin Philosophical
  Magazine and Journal of Science}\ }\textbf {\bibinfo {volume} {22}},\
  \bibinfo {pages} {983--987} (\bibinfo {year} {1936})}\BibitemShut {NoStop}%
\bibitem [{\citenamefont {Taylor}(1927)}]{taylor2002electromotive}%
  \BibitemOpen
  \bibfield  {author} {\bibinfo {author} {\bibfnamefont {P.~B.}\ \bibnamefont
  {Taylor}},\ }\bibfield  {title} {\enquote {\bibinfo {title} {Electromotive
  force of the cell with transference and theory of interdiffusion of
  electrolytes},}\ }\href@noop {} {\bibfield  {journal} {\bibinfo  {journal}
  {The Journal of Physical Chemistry}\ }\textbf {\bibinfo {volume} {31}},\
  \bibinfo {pages} {1478--1500} (\bibinfo {year} {1927})}\BibitemShut {NoStop}%
\bibitem [{\citenamefont {Gibbs}(1961)}]{gibbs1961reprinted}%
  \BibitemOpen
  \bibfield  {author} {\bibinfo {author} {\bibfnamefont {J.}~\bibnamefont
  {Gibbs}},\ }\href@noop {} {\emph {\bibinfo {title} {The Scientiﬁc Papers of
  J. Willard Gibbs}}},\ Vol.~\bibinfo {volume} {1}\ (\bibinfo  {publisher}
  {Dover Publications, New York},\ \bibinfo {year} {1961})\ p.\ \bibinfo
  {pages} {429}\BibitemShut {NoStop}%
\bibitem [{\citenamefont {Guggenheim}(1929)}]{GuggenheimI}%
  \BibitemOpen
  \bibfield  {author} {\bibinfo {author} {\bibfnamefont {E.~A.}\ \bibnamefont
  {Guggenheim}},\ }\bibfield  {title} {\enquote {\bibinfo {title} {The
  conceptions of electrical potential difference between two phases and the
  individual activities of ions},}\ }\href@noop {} {\bibfield  {journal}
  {\bibinfo  {journal} {The Journal of Physical Chemistry}\ }\textbf {\bibinfo
  {volume} {33}},\ \bibinfo {pages} {842--849} (\bibinfo {year}
  {1929})}\BibitemShut {NoStop}%
\bibitem [{\citenamefont {Guggenheim}(1930)}]{GuggenheimII}%
  \BibitemOpen
  \bibfield  {author} {\bibinfo {author} {\bibfnamefont {E.~A.}\ \bibnamefont
  {Guggenheim}},\ }\bibfield  {title} {\enquote {\bibinfo {title} {On the
  conception of electrical potential difference between two phases. ii},}\
  }\href@noop {} {\bibfield  {journal} {\bibinfo  {journal} {The Journal of
  Physical Chemistry}\ }\textbf {\bibinfo {volume} {34}},\ \bibinfo {pages}
  {1540--1543} (\bibinfo {year} {1930})}\BibitemShut {NoStop}%
\bibitem [{\citenamefont {Labbez}\ and\ \citenamefont
  {J{\"o}nsson}(2006)}]{labbez2006new}%
  \BibitemOpen
  \bibfield  {author} {\bibinfo {author} {\bibfnamefont {C.}~\bibnamefont
  {Labbez}}\ and\ \bibinfo {author} {\bibfnamefont {B.}~\bibnamefont
  {J{\"o}nsson}},\ }\href@noop {} {\emph {\bibinfo {title} {A new Monte Carlo
  method for the titration of molecules and minerals}}}\ (\bibinfo  {publisher}
  {Springer},\ \bibinfo {year} {2006})\ pp.\ \bibinfo {pages}
  {66--72}\BibitemShut {NoStop}%
\bibitem [{\citenamefont {Bakhshandeh}, \citenamefont {Frydel},\ and\
  \citenamefont {Levin}(2022{\natexlab{a}})}]{bakhshandeh2022reactive}%
  \BibitemOpen
  \bibfield  {author} {\bibinfo {author} {\bibfnamefont {A.}~\bibnamefont
  {Bakhshandeh}}, \bibinfo {author} {\bibfnamefont {D.}~\bibnamefont {Frydel}},
  \ and\ \bibinfo {author} {\bibfnamefont {Y.}~\bibnamefont {Levin}},\
  }\bibfield  {title} {\enquote {\bibinfo {title} {Reactive monte carlo
  simulations for charge regulation of colloidal particles},}\ }\href@noop {}
  {\bibfield  {journal} {\bibinfo  {journal} {The Journal of Chemical Physics}\
  }\textbf {\bibinfo {volume} {156}} (\bibinfo {year}
  {2022}{\natexlab{a}})}\BibitemShut {NoStop}%
\bibitem [{\citenamefont {Levin}\ and\ \citenamefont
  {Bakhshandeh}(2023{\natexlab{a}})}]{comm}%
  \BibitemOpen
  \bibfield  {author} {\bibinfo {author} {\bibfnamefont {Y.}~\bibnamefont
  {Levin}}\ and\ \bibinfo {author} {\bibfnamefont {A.}~\bibnamefont
  {Bakhshandeh}},\ }\bibfield  {title} {\enquote {\bibinfo {title} {{A new
  method for reactive constant pH simulations}},}\ }\href {\doibase
  10.1063/5.0166840} {\bibfield  {journal} {\bibinfo  {journal} {The Journal of
  Chemical Physics}\ }\textbf {\bibinfo {volume} {159}},\ \bibinfo {pages}
  {111101} (\bibinfo {year} {2023}{\natexlab{a}})},\ \Eprint
  {http://arxiv.org/abs/https://pubs.aip.org/aip/jcp/article-pdf/doi/10.1063/5.0166840/18128194/111101\_1\_5.0166840.pdf}
  {https://pubs.aip.org/aip/jcp/article-pdf/doi/10.1063/5.0166840/18128194/111101\_1\_5.0166840.pdf}
  \BibitemShut {NoStop}%
\bibitem [{\citenamefont {Craxford}, \citenamefont {Gatty},\ and\ \citenamefont
  {Teorell}(1938)}]{craxford1938xcv}%
  \BibitemOpen
  \bibfield  {author} {\bibinfo {author} {\bibfnamefont {S.}~\bibnamefont
  {Craxford}}, \bibinfo {author} {\bibfnamefont {O.}~\bibnamefont {Gatty}}, \
  and\ \bibinfo {author} {\bibfnamefont {T.}~\bibnamefont {Teorell}},\
  }\bibfield  {title} {\enquote {\bibinfo {title} {Xcv. a note on surface
  ph},}\ }\href@noop {} {\bibfield  {journal} {\bibinfo  {journal} {The London,
  Edinburgh, and Dublin Philosophical Magazine and Journal of Science}\
  }\textbf {\bibinfo {volume} {25}},\ \bibinfo {pages} {1061--1066} (\bibinfo
  {year} {1938})}\BibitemShut {NoStop}%
\bibitem [{\citenamefont {van Roij}(2003)}]{van2003defying}%
  \BibitemOpen
  \bibfield  {author} {\bibinfo {author} {\bibfnamefont {R.}~\bibnamefont {van
  Roij}},\ }\bibfield  {title} {\enquote {\bibinfo {title} {Defying gravity
  with entropy and electrostatics: sedimentation of charged colloids},}\
  }\href@noop {} {\bibfield  {journal} {\bibinfo  {journal} {Journal of
  Physics: Condensed Matter}\ }\textbf {\bibinfo {volume} {15}},\ \bibinfo
  {pages} {S3569} (\bibinfo {year} {2003})}\BibitemShut {NoStop}%
\bibitem [{\citenamefont {Philipse}(2004)}]{philipse2004remarks}%
  \BibitemOpen
  \bibfield  {author} {\bibinfo {author} {\bibfnamefont {A.~P.}\ \bibnamefont
  {Philipse}},\ }\bibfield  {title} {\enquote {\bibinfo {title} {Remarks on the
  donnan condenser in the sedimentation--diffusion equilibrium of charged
  colloids},}\ }\href@noop {} {\bibfield  {journal} {\bibinfo  {journal}
  {Journal of Physics: Condensed Matter}\ }\textbf {\bibinfo {volume} {16}},\
  \bibinfo {pages} {S4051} (\bibinfo {year} {2004})}\BibitemShut {NoStop}%
\bibitem [{\citenamefont {Warren}(2023)}]{warren2004electrifying}%
  \BibitemOpen
  \bibfield  {author} {\bibinfo {author} {\bibfnamefont {P.~B.}\ \bibnamefont
  {Warren}},\ }\bibfield  {title} {\enquote {\bibinfo {title} {Partial osmotic
  pressures of ions in electrolyte solutions and the gibbs-guggenheim
  uncertainty principle},}\ }\href@noop {} {\bibfield  {journal} {\bibinfo
  {journal} {Physical Review E}\ }\textbf {\bibinfo {volume} {107}},\ \bibinfo
  {pages} {034606} (\bibinfo {year} {2023})}\BibitemShut {NoStop}%
\bibitem [{\citenamefont {Brezinski}(1983)}]{BREZINSKI1983347}%
  \BibitemOpen
  \bibfield  {author} {\bibinfo {author} {\bibfnamefont {D.~P.}\ \bibnamefont
  {Brezinski}},\ }\bibfield  {title} {\enquote {\bibinfo {title} {Influence of
  colloidal charge on response of ph and reference electrodes: the suspension
  effect},}\ }\href {\doibase https://doi.org/10.1016/0039-9140(83)80078-2}
  {\bibfield  {journal} {\bibinfo  {journal} {Talanta}\ }\textbf {\bibinfo
  {volume} {30}},\ \bibinfo {pages} {347--354} (\bibinfo {year}
  {1983})}\BibitemShut {NoStop}%
\bibitem [{\citenamefont {Barr}\ and\ \citenamefont
  {Panagiotopoulos}(2012)}]{barr2012grand}%
  \BibitemOpen
  \bibfield  {author} {\bibinfo {author} {\bibfnamefont {S.}~\bibnamefont
  {Barr}}\ and\ \bibinfo {author} {\bibfnamefont {A.}~\bibnamefont
  {Panagiotopoulos}},\ }\bibfield  {title} {\enquote {\bibinfo {title}
  {Grand-canonical monte carlo method for donnan equilibria},}\ }\href@noop {}
  {\bibfield  {journal} {\bibinfo  {journal} {Physical Review E}\ }\textbf
  {\bibinfo {volume} {86}},\ \bibinfo {pages} {016703} (\bibinfo {year}
  {2012})}\BibitemShut {NoStop}%
\bibitem [{\citenamefont {Levin}\ and\ \citenamefont
  {Bakhshandeh}(2023{\natexlab{b}})}]{Levincomment}%
  \BibitemOpen
  \bibfield  {author} {\bibinfo {author} {\bibfnamefont {Y.}~\bibnamefont
  {Levin}}\ and\ \bibinfo {author} {\bibfnamefont {A.}~\bibnamefont
  {Bakhshandeh}},\ }\bibfield  {title} {\enquote {\bibinfo {title} {Comment on
  “simulations of ionization equilibria in weak polyelectrolyte solutions and
  gels” by j. landsgesell{,} l. nová{,} o. rud{,} f. uhlík{,} d. sean{,} p.
  hebbeker{,} c. holm and p. košovan{,} soft matter{,} 2019{,} 15{,}
  1155–1185},}\ }\href@noop {} {\bibfield  {journal} {\bibinfo  {journal}
  {Soft Matter}\ }\textbf {\bibinfo {volume} {19}},\ \bibinfo {pages}
  {3519--3521} (\bibinfo {year} {2023}{\natexlab{b}})}\BibitemShut {NoStop}%
\bibitem [{\citenamefont {Landsgesell}\ \emph {et~al.}(2019)\citenamefont
  {Landsgesell}, \citenamefont {Nov{\'a}}, \citenamefont {Rud}, \citenamefont
  {Uhl{\'\i}k}, \citenamefont {Sean}, \citenamefont {Hebbeker}, \citenamefont
  {Holm},\ and\ \citenamefont {Ko{\v{s}}ovan}}]{landsgesell2019simulations}%
  \BibitemOpen
  \bibfield  {author} {\bibinfo {author} {\bibfnamefont {J.}~\bibnamefont
  {Landsgesell}}, \bibinfo {author} {\bibfnamefont {L.}~\bibnamefont
  {Nov{\'a}}}, \bibinfo {author} {\bibfnamefont {O.}~\bibnamefont {Rud}},
  \bibinfo {author} {\bibfnamefont {F.}~\bibnamefont {Uhl{\'\i}k}}, \bibinfo
  {author} {\bibfnamefont {D.}~\bibnamefont {Sean}}, \bibinfo {author}
  {\bibfnamefont {P.}~\bibnamefont {Hebbeker}}, \bibinfo {author}
  {\bibfnamefont {C.}~\bibnamefont {Holm}}, \ and\ \bibinfo {author}
  {\bibfnamefont {P.}~\bibnamefont {Ko{\v{s}}ovan}},\ }\bibfield  {title}
  {\enquote {\bibinfo {title} {Simulations of ionization equilibria in weak
  polyelectrolyte solutions and gels},}\ }\href@noop {} {\bibfield  {journal}
  {\bibinfo  {journal} {Soft Matter}\ }\textbf {\bibinfo {volume} {15}},\
  \bibinfo {pages} {1155--1185} (\bibinfo {year} {2019})}\BibitemShut {NoStop}%
\bibitem [{\citenamefont {Bakhshandeh}, \citenamefont {Frydel},\ and\
  \citenamefont {Levin}(2022{\natexlab{b}})}]{bkh22}%
  \BibitemOpen
  \bibfield  {author} {\bibinfo {author} {\bibfnamefont {A.}~\bibnamefont
  {Bakhshandeh}}, \bibinfo {author} {\bibfnamefont {D.}~\bibnamefont {Frydel}},
  \ and\ \bibinfo {author} {\bibfnamefont {Y.}~\bibnamefont {Levin}},\
  }\bibfield  {title} {\enquote {\bibinfo {title} {Theory of charge regulation
  of colloidal particles in electrolyte solutions},}\ }\href@noop {} {\bibfield
   {journal} {\bibinfo  {journal} {Langmuir}\ }\textbf {\bibinfo {volume}
  {38}},\ \bibinfo {pages} {13963--13971} (\bibinfo {year}
  {2022}{\natexlab{b}})}\BibitemShut {NoStop}%
\bibitem [{\citenamefont {Fuoss}, \citenamefont {Katchalsky},\ and\
  \citenamefont {Lifson}(1951)}]{fuoss1951potential}%
  \BibitemOpen
  \bibfield  {author} {\bibinfo {author} {\bibfnamefont {R.~M.}\ \bibnamefont
  {Fuoss}}, \bibinfo {author} {\bibfnamefont {A.}~\bibnamefont {Katchalsky}}, \
  and\ \bibinfo {author} {\bibfnamefont {S.}~\bibnamefont {Lifson}},\
  }\bibfield  {title} {\enquote {\bibinfo {title} {The potential of an infinite
  rod-like molecule and the distribution of the counter ions},}\ }\href@noop {}
  {\bibfield  {journal} {\bibinfo  {journal} {Proceedings of the National
  Academy of Sciences}\ }\textbf {\bibinfo {volume} {37}},\ \bibinfo {pages}
  {579--589} (\bibinfo {year} {1951})}\BibitemShut {NoStop}%
\bibitem [{\citenamefont {Marcus}(1955)}]{marcus1955calculation}%
  \BibitemOpen
  \bibfield  {author} {\bibinfo {author} {\bibfnamefont {R.}~\bibnamefont
  {Marcus}},\ }\bibfield  {title} {\enquote {\bibinfo {title} {Calculation of
  thermodynamic properties of polyelectrolytes},}\ }\href@noop {} {\bibfield
  {journal} {\bibinfo  {journal} {The Journal of Chemical Physics}\ }\textbf
  {\bibinfo {volume} {23}},\ \bibinfo {pages} {1057--1068} (\bibinfo {year}
  {1955})}\BibitemShut {NoStop}%
\bibitem [{\citenamefont {Carnahan}\ and\ \citenamefont
  {Starling}(1969)}]{carnahan1969equation}%
  \BibitemOpen
  \bibfield  {author} {\bibinfo {author} {\bibfnamefont {N.~F.}\ \bibnamefont
  {Carnahan}}\ and\ \bibinfo {author} {\bibfnamefont {K.~E.}\ \bibnamefont
  {Starling}},\ }\bibfield  {title} {\enquote {\bibinfo {title} {Equation of
  state for nonattracting rigid spheres},}\ }\href@noop {} {\bibfield
  {journal} {\bibinfo  {journal} {The Journal of Chemical Physics}\ }\textbf
  {\bibinfo {volume} {51}},\ \bibinfo {pages} {635--636} (\bibinfo {year}
  {1969})}\BibitemShut {NoStop}%
\bibitem [{\citenamefont {Carnahan}\ and\ \citenamefont
  {Starling}(1970)}]{carnahan1970thermodynamic}%
  \BibitemOpen
  \bibfield  {author} {\bibinfo {author} {\bibfnamefont {N.~F.}\ \bibnamefont
  {Carnahan}}\ and\ \bibinfo {author} {\bibfnamefont {K.~E.}\ \bibnamefont
  {Starling}},\ }\bibfield  {title} {\enquote {\bibinfo {title} {Thermodynamic
  properties of a rigid-sphere fluid},}\ }\href@noop {} {\bibfield  {journal}
  {\bibinfo  {journal} {The Journal of Chemical Physics}\ }\textbf {\bibinfo
  {volume} {53}},\ \bibinfo {pages} {600--603} (\bibinfo {year}
  {1970})}\BibitemShut {NoStop}%
\bibitem [{\citenamefont {Adams}(1974)}]{adams1974chemical}%
  \BibitemOpen
  \bibfield  {author} {\bibinfo {author} {\bibfnamefont {D.}~\bibnamefont
  {Adams}},\ }\bibfield  {title} {\enquote {\bibinfo {title} {Chemical
  potential of hard-sphere fluids by monte carlo methods},}\ }\href@noop {}
  {\bibfield  {journal} {\bibinfo  {journal} {Molecular Physics}\ }\textbf
  {\bibinfo {volume} {28}},\ \bibinfo {pages} {1241--1252} (\bibinfo {year}
  {1974})}\BibitemShut {NoStop}%
\bibitem [{\citenamefont {Maciel}, \citenamefont {Abreu},\ and\ \citenamefont
  {Tavares}(2018)}]{MACIEL2018}%
  \BibitemOpen
  \bibfield  {author} {\bibinfo {author} {\bibfnamefont {J.~A. C. d. S.~L.}\
  \bibnamefont {Maciel}}, \bibinfo {author} {\bibfnamefont {C.~R.~A.}\
  \bibnamefont {Abreu}}, \ and\ \bibinfo {author} {\bibfnamefont {F.~W.}\
  \bibnamefont {Tavares}},\ }\bibfield  {title} {\enquote {\bibinfo {title}
  {Chemical potentials of hard-core molecules bt a stepwise insertion
  method},}\ }\href@noop {} {\bibfield  {journal} {\bibinfo  {journal}
  {Brazilian Journal of Chemical Engineering}\ }\textbf {\bibinfo {volume}
  {35}},\ \bibinfo {pages} {277--288} (\bibinfo {year} {2018})}\BibitemShut
  {NoStop}%
\bibitem [{\citenamefont {H{\o}ye}\ and\ \citenamefont
  {Lomba}(1988)}]{ho1988mean}%
  \BibitemOpen
  \bibfield  {author} {\bibinfo {author} {\bibfnamefont {J.~S.}\ \bibnamefont
  {H{\o}ye}}\ and\ \bibinfo {author} {\bibfnamefont {E.}~\bibnamefont
  {Lomba}},\ }\bibfield  {title} {\enquote {\bibinfo {title} {Mean spherical
  approximation (msa) for a simple model of electrolytes. i. theoretical
  foundations and thermodynamics},}\ }\href@noop {} {\bibfield  {journal}
  {\bibinfo  {journal} {The Journal of Chemical Physics}\ }\textbf {\bibinfo
  {volume} {88}},\ \bibinfo {pages} {5790--5797} (\bibinfo {year}
  {1988})}\BibitemShut {NoStop}%
\bibitem [{\citenamefont {Ho}, \citenamefont {Tsao},\ and\ \citenamefont
  {Sheng}(2003)}]{ho2003interfacial}%
  \BibitemOpen
  \bibfield  {author} {\bibinfo {author} {\bibfnamefont {C.-H.}\ \bibnamefont
  {Ho}}, \bibinfo {author} {\bibfnamefont {H.-K.}\ \bibnamefont {Tsao}}, \ and\
  \bibinfo {author} {\bibfnamefont {Y.-J.}\ \bibnamefont {Sheng}},\ }\bibfield
  {title} {\enquote {\bibinfo {title} {Interfacial tension of a salty droplet:
  Monte carlo study},}\ }\href@noop {} {\bibfield  {journal} {\bibinfo
  {journal} {The Journal of Chemical Physics}\ }\textbf {\bibinfo {volume}
  {119}},\ \bibinfo {pages} {2369--2375} (\bibinfo {year} {2003})}\BibitemShut
  {NoStop}%
\bibitem [{\citenamefont {Levin}\ and\ \citenamefont
  {Fisher}(1996)}]{levin1996criticality}%
  \BibitemOpen
  \bibfield  {author} {\bibinfo {author} {\bibfnamefont {Y.}~\bibnamefont
  {Levin}}\ and\ \bibinfo {author} {\bibfnamefont {M.~E.}\ \bibnamefont
  {Fisher}},\ }\bibfield  {title} {\enquote {\bibinfo {title} {Criticality in
  the hard-sphere ionic fluid},}\ }\href@noop {} {\bibfield  {journal}
  {\bibinfo  {journal} {Physica A: Statistical Mechanics and its Applications}\
  }\textbf {\bibinfo {volume} {225}},\ \bibinfo {pages} {164--220} (\bibinfo
  {year} {1996})}\BibitemShut {NoStop}%
\bibitem [{\citenamefont {Waisman}\ and\ \citenamefont
  {Lebowitz}(1972)}]{waisman1972mean}%
  \BibitemOpen
  \bibfield  {author} {\bibinfo {author} {\bibfnamefont {E.}~\bibnamefont
  {Waisman}}\ and\ \bibinfo {author} {\bibfnamefont {J.~L.}\ \bibnamefont
  {Lebowitz}},\ }\bibfield  {title} {\enquote {\bibinfo {title} {Mean spherical
  model integral equation for charged hard spheres i. method of solution},}\
  }\href@noop {} {\bibfield  {journal} {\bibinfo  {journal} {The Journal of
  Chemical Physics}\ }\textbf {\bibinfo {volume} {56}},\ \bibinfo {pages}
  {3086--3093} (\bibinfo {year} {1972})}\BibitemShut {NoStop}%
\bibitem [{\citenamefont {Blum}(1975)}]{blum1975mean}%
  \BibitemOpen
  \bibfield  {author} {\bibinfo {author} {\bibfnamefont {L.}~\bibnamefont
  {Blum}},\ }\bibfield  {title} {\enquote {\bibinfo {title} {Mean spherical
  model for asymmetric electrolytes: I. method of solution},}\ }\href@noop {}
  {\bibfield  {journal} {\bibinfo  {journal} {Molecular Physics}\ }\textbf
  {\bibinfo {volume} {30}},\ \bibinfo {pages} {1529--1535} (\bibinfo {year}
  {1975})}\BibitemShut {NoStop}%
\bibitem [{\citenamefont {Levin}(2002)}]{levin2002electrostatic}%
  \BibitemOpen
  \bibfield  {author} {\bibinfo {author} {\bibfnamefont {Y.}~\bibnamefont
  {Levin}},\ }\bibfield  {title} {\enquote {\bibinfo {title} {Electrostatic
  correlations: from plasma to biology},}\ }\href@noop {} {\bibfield  {journal}
  {\bibinfo  {journal} {Reports on Progress in Physics}\ }\textbf {\bibinfo
  {volume} {65}},\ \bibinfo {pages} {1577} (\bibinfo {year}
  {2002})}\BibitemShut {NoStop}%
\bibitem [{\citenamefont {Bakhshandeh}\ \emph {et~al.}(2019)\citenamefont
  {Bakhshandeh}, \citenamefont {Frydel}, \citenamefont {Diehl},\ and\
  \citenamefont {Levin}}]{bakhshandeh2019}%
  \BibitemOpen
  \bibfield  {author} {\bibinfo {author} {\bibfnamefont {A.}~\bibnamefont
  {Bakhshandeh}}, \bibinfo {author} {\bibfnamefont {D.}~\bibnamefont {Frydel}},
  \bibinfo {author} {\bibfnamefont {A.}~\bibnamefont {Diehl}}, \ and\ \bibinfo
  {author} {\bibfnamefont {Y.}~\bibnamefont {Levin}},\ }\bibfield  {title}
  {\enquote {\bibinfo {title} {Charge regulation of colloidal particles: Theory
  and simulations},}\ }\href@noop {} {\bibfield  {journal} {\bibinfo  {journal}
  {Physical Review Letters}\ }\textbf {\bibinfo {volume} {123}},\ \bibinfo
  {pages} {208004} (\bibinfo {year} {2019})}\BibitemShut {NoStop}%
\bibitem [{\citenamefont {Bjerrum}(1958)}]{bjerrum1958electrical}%
  \BibitemOpen
  \bibfield  {author} {\bibinfo {author} {\bibfnamefont {N.}~\bibnamefont
  {Bjerrum}},\ }\bibfield  {title} {\enquote {\bibinfo {title} {The electrical
  contact potential and the individual activity},}\ }\href@noop {} {\bibfield
  {journal} {\bibinfo  {journal} {Acta Chem. Scand}\ }\textbf {\bibinfo
  {volume} {12}} (\bibinfo {year} {1958})}\BibitemShut {NoStop}%
\bibitem [{Note1()}]{Note1}%
  \BibitemOpen
  \bibinfo {note} {For reasons why it is not possible to uniquely define the
  ``mean}\BibitemShut {NoStop}%
\end{thebibliography}%


	\end{document}